\documentclass[final]{siamltex}

\usepackage{epsfig, graphicx}
\usepackage{caption}

\begin{document}

\title{Data for Development:\\ The D4D Challenge on Mobile Phone Data}
\author{
Vincent  D. Blondel%
\thanks{University of Louvain, B-1348 Louvain-la-Neuve, Belgium. {\tt vincent.blondel@uclouvain.be}}
\and Markus Esch%
  \footnotemark[1]
  \and Connie Chan%
    \footnotemark[1]
  \and 
Fabrice Clerot%
\thanks{Orange Labs, France}
\and Pierre Deville%
  \footnotemark[1]
\and Etienne Huens%
  \footnotemark[1]
\and Fr\'ed\'eric Morlot%
   \footnotemark[2]
\and Zbigniew Smoreda%
   \footnotemark[2]
\and Cezary Ziemlicki%
   \footnotemark[2]
  }
  
  \date{}
\maketitle

\begin{abstract}        
The Orange ``Data for Development" (D4D) challenge is an open data challenge on anonymous call patterns of Orange's mobile phone users in Ivory Coast. The goal of the challenge is to help address society development questions in novel ways by contributing to the socio-economic development and well-being of the Ivory Coast population. Participants to the challenge are given access to four mobile phone datasets and the purpose of this paper is to describe the four datasets. The website {\tt http://www.d4d.orange.com} contains more information about the participation rules. 

 The datasets are based on anonymized Call Detail Records (CDR) of phone calls and SMS exchanges between five million of Orange's customers in Ivory Coast between December 1, 2011 and April 28, 2012. The datasets are: (a) antenna-to-antenna traffic on an hourly basis, (b) individual trajectories for 50,000 customers for two week time windows with antenna location information, (3) individual trajectories for 500,000 customers over the entire observation period with sub-prefecture location information, and (4) a sample of communication graphs for  5,000 customers.
\end{abstract}


\begin{figure}[!ht]
	\centering
		\includegraphics[width=0.9\columnwidth]{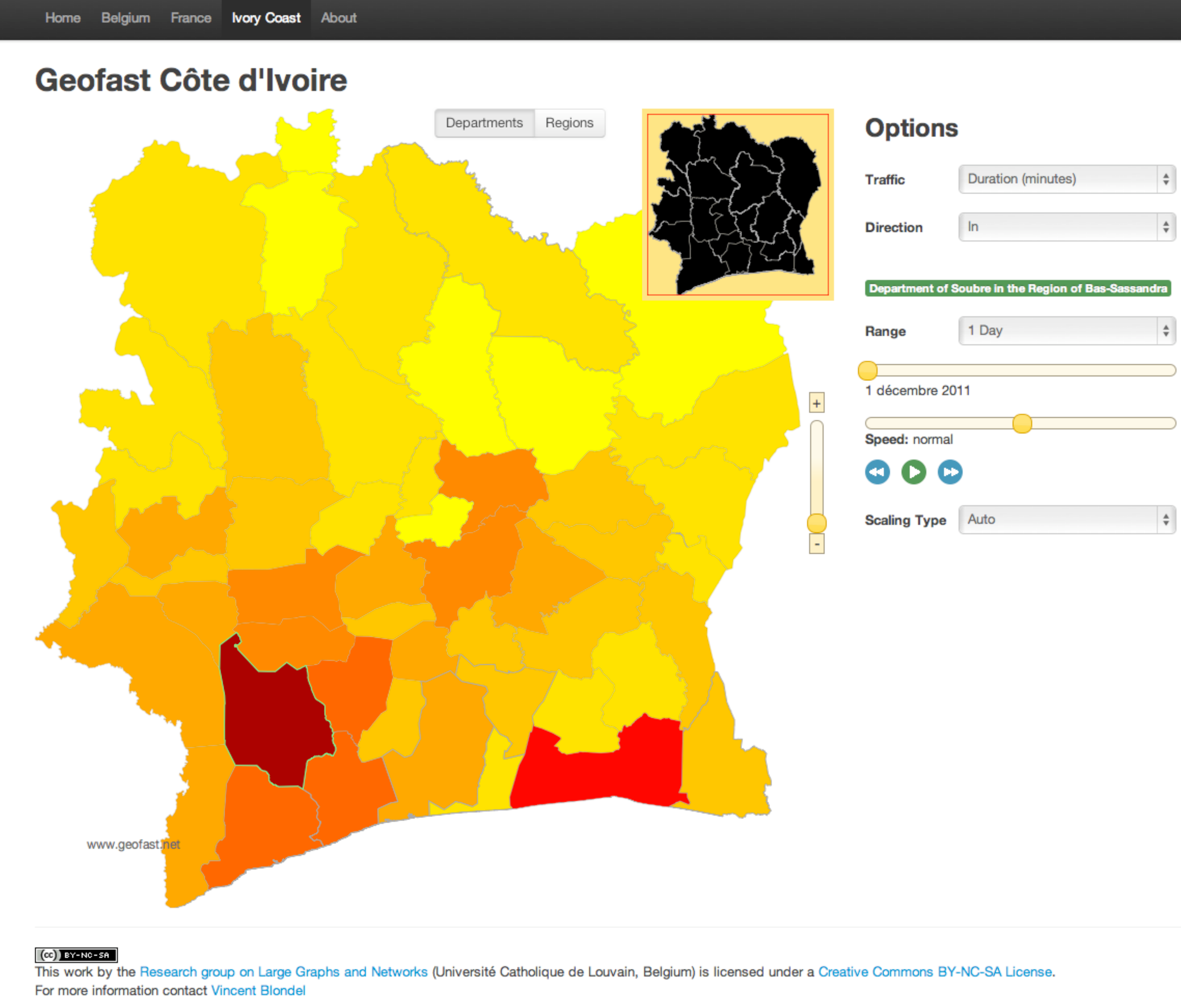}
	\caption*{The geofast web interface {\tt www.geofast.net} for the visualisation of mobile phone communications (countries available: France, Belgium, Ivory Coast).}
	\label{geo}	
\end{figure}

\newpage

\section{Introduction}
The availability of detailed mobility traces and mobile phone communication data for large populations has already had a significant impact on research in behavioral science. Some researchers consider such datasets as an opportunity to refine  the analysis of human behavior \cite{La}, while others question the usefulness of such  datasets to draw conclusions on collective human behavior \cite{BB,BC,La}.

Digital traces left by mobile phone users often reveal sensitive private individual information. It is therefore natural to limit access to such data. Limited access to data of scientific interest is however a potential source of  a ``new digital divide" in the scientific community, as described in \cite{BC}. In order to improve the availability of large mobile phone datasets and to foster research in this area, the Orange Group decided to provide anonymized datasets from Ivory Coast for the purpose of scientific research. With around five million customers, Orange has a significant market share in Ivory Coast, whose total population is estimated to be 20 million individuals. In addition to the scientific benefit, the project intends to foster development in Ivory Coast by establishing new collaborations with African scientists and by providing behavioral data that has not yet been collected by the national statistics agency \cite{E}.

\section{Other datasets on Ivory Coast}

Researchers participating in the D4D challenge are encouraged to combine the D4D mobile phone datasets with other datasets and source of information. These sources include the following.\\

African Development Bank Group. The African Development Bank (AfDB) GroupÕs mission is to help reduce poverty, improve living conditions for Africans and mobilize resources for the continentÕs economic and social development. With this objective in mind, the institution aims at assisting African countries in their efforts to achieve sustainable economic development and social progress.\\
{\tt http://www.afdb.org/en/}\\

African Economic Outlook. Economic, social and political developments of African countries, with the expertise of the African Development Bank, the OECD Development Centre, the United Nations Economic Commission for Africa, the United Nations Development Programme and a network of African think tanks and research centres.\\
http://www.africaneconomicoutlook.org/en/\\

Africa and Middle East Telecom News. Africa and Middle East Telecom-Week' tracks the fixed, broadband and mobile phone markets in Africa and Middle East.\\
{\tt http://www.africantelecomsnews.com/}\\

Africa Renewal on Line. The Africa Renewal magazine is produced by the United Nations organism and provides up-to-date information and analysis of the major economic and development challenges facing Africa today. It works with the media in Africa and beyond to promote the work of the United Nations, Africa and the international community to bring peace and development to Africa.\\
{\tt http://www.un.org/french/ecosocdev/geninfo/afrec/vol25n}\\

Africa Research Program. Data set consists of an aggregate of a number of the most commonly used publicly available variables used in the study of African political economy.\\
{\tt http://africa.gov.harvard.edu/}\\

African Union. Pan African Organization.\\
{\tt http://au.int/en/resources/documents}\\

Africover. Geographic data produced by the the Africover Project and the participating countries. The information available in the national Multipurpose Africover Databases on Environmental Resources (MADE) is composed by a main geographic information layer (i.e. land cover) and several additional layers that vary for each country (e.g. roads, rivers and water bodies, etc.); the available data produced by Africover is listed for each country in the Africover Data table (here only full resolution data sets and public domain spatially aggregated data sets are listed. Thematic aggregations are available starting from the metadata of these data sets).\\
{\tt http://www.africover.org/}\\

Afristat. Observatoire Economique et Statistique dÕAfrique Subsaharienne\\
{\tt http://www.afristat.org/publication/acces-direct-aux-donnees}\\

Banque Centrale des Etats de l'Afrique de l'Ouest. Financial and economic data.
{\tt http://edenpub.bceao.int/}\\

Center For International Development. This page is a depository for data developed through research at the Center for International Development at Harvard University (CID). Often the data are associated with a research paper and, thus, the paper is also available for downloading.\\
{\tt http://www.cid.harvard.edu/ciddata/ciddata.html}\\

CIA - The World Factbook. The World Factbook provides information on the history, people, government, economy, geography, communications, transportation, military, and transnational issues for 267 world entities.\\
{\tt https://www.cia.gov/library/publications/the-world-factbook/}\\

Factset: A compilation of various international economic data sets.\\
{\tt http://www.factset.com}\\

Famine Early Warning Systems Network. The Famine Early Warning Systems Network (FEWS NET) is a US AID-funded activity that collaborates with international, regional and national partners to provide timely and rigorous early warning and vulnerability information on emerging and evolving food security issues.\\
{\tt http://www.fews.net/Pages/}\\

Food and Agriculture Organization of the United Nations. FAO's mandate is to raise levels of nutrition, improve agricultural productivity, better the lives of rural populations and contribute to the growth of the world economy.\\
{\tt http://www.fao.org/corp/statistics/en/}\\

Global Distribution of Poverty. A website with a collection of subnational, spatially explicit, poverty data sets. This page is maintained by The Poverty Mapping Project at CIESIN (The Center for International Earth Science Information Network) at the Earth Institute at Columbia University.\\
{\tt http://sedac.ciesin.columbia.edu/povmap/}\\

International Census. Global population trends, links to historical population estimates, population clocks, and estimates of population, births, and deaths occurring each year, day, hour, or second.\\
{\tt http://www.census.gov/ipc/www/idb/}\\

Investir en zone France. Economic data about african french-speaking countries (glossary, economic indicators, maps, etc.).\\
{\tt http://www.izf.net/bdd-entreprise/}\\

ITU - Telecommunication Development Sector.\\
{\tt http://www.itu.int/net/ITU-D/}\\

Measure DHS. Information about population, health and nutrition programs.\\
{\tt http://www.measuredhs.com/}\\

Measuring the Information Society. ICT Indicators for Development. ICT measurement is a tool for policymakers, to assess the status of ICT in developing countries and craft policies to maximize the benefits of ICT for those countries.\\
{\tt http://new.unctad.org/}\\

Princeton Data and Statistical Services: Data on Africa. A compilation of datasets on Africa.\\
{\tt http://dss1.princeton.edu/cgi-bin/dataresources/newdataresources.cgi?term=14}\\

Research ICT Africa Network. The Research ICT Africa Network conducts research on ICT policy and regulation that facilitates evidence-based and informed policy making for improved access, use and application of ICT for social development and economic growth.\\
{\tt http://www.researchictafrica.net/home.php}\\

The International Aid Transparency Initiative. The International Aid Transparency Initiative aims to make information about aid spending easier to access, use and understand.\\
{\tt http://www.aidtransparency.net/}\\

United Nations Data. The United Nations Statistics Division (UNSD) launched a new internet based data service for the global user community. It brings UN statistical databases within easy reach of users through a single entry point.\\
{\tt http://data.un.org/}\\

United Nations Economic Commission for Africa. ECA's mandate is to promote the economic and social development of its member States, foster intra-regional integration, and promote international cooperation for Africa's development.\\
{\tt http://www.uneca.org/}\\

US Census International Bureau Programs. The U.S. Census Bureau conducts demographic, economic, and geographic studies of countries around the world.\\
{\tt http://www.census.gov/population/international/}\\

World Bank Data. The World Bank provides free and open access to a comprehensive set of data about development in countries around the globe.\\
{\tt http://data.worldbank.org/}\\

World Trade Organization. Interactive access to the most up-to-date WTO trade statistics.\\
{\tt http://www.wto.org/}\\

Mobile and Development Intelligence GSMA. MDI is an Open Data portal for the developing world mobile industry. A challenge facing mobile industry stakeholders in the developing world is the lack of publicly available data and analysis to support their business decision making and to clarify the socio-economic impact of mobile. MDI will fill this information gap and will aggregate and host data from multiple sources such as the World Bank, UN, member operators and from vendors and development organisations.\\
{\tt http://mobiledevelopmentintelligence.com/}\\

Centre sur les politiques internationales des TIC pour les pays de l'Afrique de l'Ouest. CIPACO has been initiated by Panos Institute West Africa (PIWA - a regional West African NGO), in order to strengthen the capacity of African stakeholders for an effective participation in ICT decision-making processes.\\
{\tt http://www.cipaco.org/index.php}\\

Institut National de la Statistique - R\'epublique de C\^ote d'Ivoire. General information about the country data.\\
{\tt http://www.ins.ci/}\\

\section{Data Preprocessing}
The data was collected for 150 days, from December 1, 2011 until April 28, 2012. The original set of {Call Detail Records} (CDRs) contains 2.5 billion calls and SMS exchanges between around five million users. CDRs have the following standard format: {\tt timestamp, caller\_id, callee\_id, call\_duration, antenna\_code}. The customer identifiers were anonymized by Orange Ivory Coast and all subsequent data processing was completed by Orange Labs in Paris.

 In order to have a homogeneous data sample, customers that subscribed or resigned from Orange during the observation period have  been removed. Additionally, incoming and outgoing calls have been paired in order to eliminate double counts (i.e. an incoming call for an individual is an outgoing call for the correspondent).

The provided datasets contain the geographical positions of cell phone antennas. Orange considers the exact antenna location as sensitive information and therefore the locations have been slightly blurred so as to  protect Orange's commercial interests. 

For technical reasons, the antenna identifiers are not always available. Instead of removing the corresponding communications,  the code $-1$ was given to antenna with missing identifier. This happens for a significant number of calls (about one in four).

The datasets covers a total of 3600 hours. Due to technical reasons data is sometimes missing in the datasets; missing data covers a total period of about 100 hours.

\section{Published Datasets}

All datasets are available in Tabulation Separated Values (TSV) plain text format.

\subsection{Antenna-to-antenna (SET1)}
For this dataset, the number of calls as well as the duration of calls between any pair of antennas have been aggregated hour by hour.  Calls spanning multiple time slots are considered to be in the time slot they started in. Antennas are uniquely identified by an antenna id and a geographic location. This data is available for the entire observation period. Communication between Orange customers and customers of other providers have been removed.

The antenna-to-antenna traffic data is provided in the files {\it SET1TSV\_0.TSV} to {\it SET1TSV\_9.TSV}. The 10 files each correspond to 14 days.  Each line in a TSV file provides the number of calls as well as the total duration of calls between a pair of antennas for a given hour.\\

The DDL code for this data is:\\
\verb+CREATE TABLE H_A_FLOWS (+\\
\verb+date_hour TIMESTAMP,+\\
\verb+originating_ant INTEGER,+\\
\verb+terminating_ant INTEGER,+\\
\verb+nb_voice_calls INTEGER,+\\
\verb+duration_voice_calls INTEGER+\\
\verb+);+\vspace{0.5cm}

Example of data: \\
\verb+2012-04-28 23:00:00	1236	786	2	96+\\
\verb+2012-04-28 23:00:00	1236	804	1	539+\\
\verb+2012-04-28 23:00:00	1236	867	3	1778+\\
\verb+2012-04-28 23:00:00	1236	939	1	1+\\
\verb+2012-04-28 23:00:00	1236	1020	6	108+\\
\verb+2012-04-28 23:00:00	1236	1065	1	1047+\\
\verb+2012-04-28 23:00:00	1236	1191	1	67+\\
\verb+2012-04-28 23:00:00	1236	1236	18	2212+\\
\verb+2012-04-28 23:00:00	1237	323	1	636+\\
\verb+2012-04-28 23:00:00	1237	710	1	252+\\

This first dataset can be visualized with Geofast {\tt www.geofast.net}. Geofast is a web-based tool for the interactive exploration of mobile phone data. The data is aggregated on different administrative levels and users are able to select administrative regions and visualize the amount of communication traffic on selected days. 

\subsection{Individual Trajectories: High Spatial Resolution Data (SET2)}

Individual movement trajectories can be approximated from the geographic location of the cell phone antennas during calls.  Limited knowledge of an individual's trajectory is often sufficient for identification and the individual can then be traced during the entire observation period. Two obvious solutions to reduce the possibility of identification are to reduce the spatial resolution or to publish trajectories only for limited periods of time. Since long term observation data as well as trajectories with a high spatial resolution have interesting scientific applications, two different datasets are published in order to balance privacy protection and scientific interest.

The first dataset contains high resolution trajectories of $50,000$ randomly sampled individuals over two-week periods. The second dataset contains the trajectories of $50,000$ randomly sampled individuals for the entire observation period but with reduced spatial resolution. We describe the first dataset in this section and the second dataset in the next section.

The original data has been split into consecutive two-week periods. In each time period, $50,000$ of the customers are randomly selected and are assigned anonymized identifiers. To protect privacy new random identifiers are chosen in every time period. Time stamps are rounded to the minute.

This dataset is in the archive {\bf SET2} and contains the files \emph{POS\_SAMPLE\_0.TSV} to \emph{POS\_SAMPLE\_9.TSV}.\\

The DDL code for the data is:\\
\verb+CREATE TABLE POS_SAMPLE_0(+\\
\verb+user_id INTEGER,+\\
\verb+connection_datetime TIMESTAMP,+\\
\verb+antenna_id INTEGER+\\
\verb+);+\vspace{0.5cm}

Example of data in \emph{POS\_SAMPLE\_0.TSV}:\\
\verb+437690 2011-12-10 10:51:00 980+\\
\verb+316462 2011-12-10 16:12:00 607+\\
\verb+277814 2011-12-10 20:48:00 560+\\
\verb+419518 2011-12-10 10:05:00 -1+\\
\verb+18945 2011-12-10 11:32:00 401+\\
\verb+283750 2011-12-10 10:16:00 10+\\
\verb+11813 2011-12-10 10:08:00 970+\\
\verb+92418 2011-12-10 21:08:00 -1+\\
\verb+287887 2011-12-10 09:48:00 583+\\

The coordinates of the antenna positions are given in the files \emph{ANT\_POS.TSV}. The DDL code for the data is:\\

\verb+CREATE TABLE ANT_POS(+\\
\verb+antenna_id INTEGER,+\\
\verb+longitude FLOAT,+\\
\verb+latitude FLOAT,+\\
\verb+);+\vspace{0.5cm}\\
Example of data in \emph{ANT\_POS.TSV}:\\
\verb+1	-4.143452	5.342044+\\
\verb+2	-3.913602	5.341612+\\
\verb+3	-3.967045	5.263331+\\
\verb+4	-4.070007	5.451365+\\
\verb+5	-3.496235	6.729410+\\
\verb+6	-3.485944	6.729422+\\
\verb+7	-3.981175	5.273144+\\
\verb+8 -3.911705	5.858010+\\
\verb+9 -4.014445	5.421120+\\

\subsection{Individual Trajectories: Long Term Data (SET3)}

\begin{figure}[!ht]
	\centering
		\includegraphics[width=0.6\columnwidth]{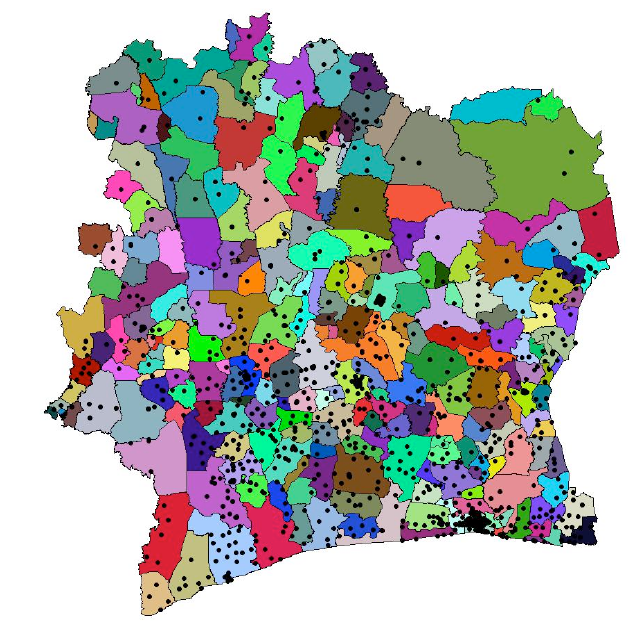}
	\caption{Orange's cell phone towers in Ivory Coast and sub-prefectures administrative regions. Note that some sub-prefectures do not have cell phone towers.}
	\label{fig:flow}	
\end{figure}

In this dataset, the trajectories of $500,000$ randomly selected individuals is provided for the entire observation period but with reduced spatial resolution. The spatial resolution is reduced by publishing the sub-prefectures of the antennas rather than the antennas' identifiers.  The published dataset also contains the geographic center of the sub-prefectures. The $255$ sub-prefectures of Ivory Coast along with Orange's cell phone towers are shown in Figure \ref{fig:flow}.

This dataset is in the archive {\bf SET3} and contains the files from  \emph{SUBPREF\_POS\_SAMPLE\_A.TSV}  to  \emph{SUBPREF\_POS\_SAMPLE\_J.TSV}, and the file \emph{SUBPREF\_POS\_LONLAT.TSV}.

The DDL code for \emph{SUBPREF\_POS\_LONLAT.TSV} is:\\

\verb+CREATE TABLE SUBPREF_POS_LONLAT(+\\
\verb+subpref_id INTEGER,+\\
\verb+longitude FLOAT,+\\
\verb+latitude FLOAT,+\\
\verb+);+\vspace{0.5cm}\\
Example of data in \emph{SUBPREF\_POS\_LONLAT.TSV}:\\
\verb+1	-3.260397	6.906417+\\
\verb+2	-3.632290	6.907771+\\
\verb+3	-3.397551	6.426104+\\
\verb+4	-3.662953	6.660800+\\
\verb+5	-3.440788	6.937723+\\
\verb+6	-3.291995	6.328551+\\
\verb+7	-3.366372	7.182663+\\
\verb+8	-3.498494	7.166416+\\
\verb+9	-3.149608	7.015214+\vspace{0.5cm}\\

The DDL code for \emph{SUBPREF\_POS\_SAMPLE.TSV} is:\\
\verb+CREATE TABLE SUBPREF_POS_SAMPLE(+\\
\verb+user_id INTEGER,+\\
\verb+connection_datetime TIMESTAMP,+\\
\verb+subpref_id INTEGER+\\
\verb+);+\vspace{0.5cm}\\

Example of data in \emph{SUBPREF\_POS\_SAMPLE\_A.TSV}:\\
\verb+134931	2011-12-02 10:50:00	60+\\
\verb+89571		2011-12-02 10:49:00	39+\\
\verb+457232	2011-12-02 16:05:00	60+\\
\verb+155864	2011-12-02 09:26:00	60+\\
\verb+280671	2011-12-02 13:24:00	-1+\\
\verb+13689		2011-12-02 20:34:00	97+\\
\verb+171642	2011-12-02 22:36:00	60+\\
\verb+247694	2011-12-02 15:11:00	60+\\
\verb+376500	2011-12-02 09:49:00	58+\\
\verb+294553	2011-12-02 20:45:00	185+\\

\subsection{Communication Subgraphs (SET4)}

\begin{figure}[!ht]
	\centering
		\includegraphics[width=0.8\columnwidth]{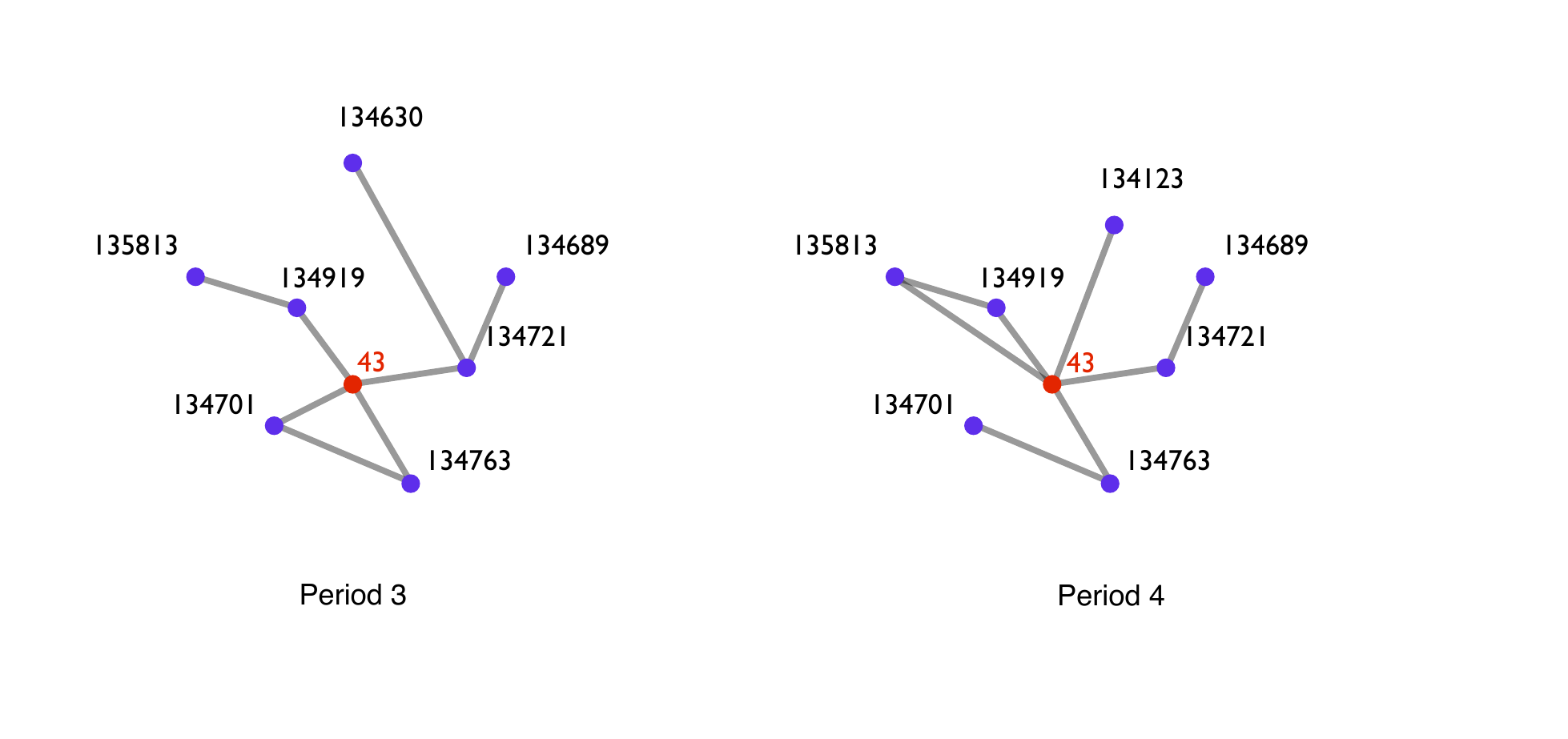}
	\caption{Ego-centered graphs. Identifiers remain unchanged during successive periods and individual appearing in two different ego-centered graphs are given different identifiers.}
	\label{geog}	
\end{figure}

Our aim with this dataset is to allow the analysis of communication graphs. The dataset contains the communication subgraphs for $5,000$ randomly selected individuals (egos). For these individuals, communications within their second order neighborhood have been divided into periods of two weeks spanning the entire observation period.  For constructing an ego-centered graph, one consider first and second order neighbors of the ego and communications between all individuals (we do however not include communications between second order neighbors). The anonymized identifiers assigned to the individuals are identical for all time slots but are unique for each subgraph. That is, a customer who is part of the communication graph of two different customers has a different identifier in the two graphs (see Figure \ref{geog}). We therefore have a total of 5,000 connected graphs in every time period. The egos have been given identifiers between 1 and 10,000 and neighbor labelling starts from 20,000.

Phone calls that follow a {public phone} usage pattern have been excluded from the randomly selected individuals. In Ivory Coast, it is common for some mobile phone owners to provide their phone to people on the street for a fee. This usage is characterized by a large number of outgoing calls but little mobility. We have removed from our selection of egos the customers identified as public phone providers.

The files are  in the archive {\bf SET4}. The communication subgraph data is published in the files \emph{GRAPHS\_0.TSV} to \emph{GRAPHS\_9.TSV}. Each of the files contains the aggregated communication graphs within the second order neighborhood of the randomly selected individuals, divided into two-week periods, starting on December 5, 2011. For every pair of individuals we indicate if there has been a communication between the two, we do not provided the number of communications, total communication time or the direction of the communication.\\

The DDL code for those data is:\\
\verb+CREATE TABLE GRAPHS_0(+\\
\verb+source_user_id INTEGER,+\\
\verb+destination_user_id INTEGER,+\\
\verb+);+\vspace{0.5cm}\\

Example of data in \emph{GRAPHS\_0.TSV}:\\
\verb+1052  20002+\\
\verb+20002 20022+\\
\verb+20018 20019+\\
\verb+1052  20019+\\
\verb+20019 20030+\\
\verb+20019 20031+\\
\verb+20129 20119+\\
\verb+20132 20119+\\
\verb+20134 20119+\\
\verb+20102 20135+\\


\end{document}